# A New Cryptosystem Based on Positive Braids

1st Xiaoming Chen
*Beijing Electronic Science and Technology*
*University of Science and Technology of China*
Beijing, China
chenxmphd@besti.edu.cn

2nd Weiqing You*
*Beijing Electronic Science and Technology*
Beijing, China
scipaperyou@sina.com

3rd Meng Jiao
*Beijing Electronic Science and Technology*
Beijing, China
jmgl8267@sina.com

4th Kejun Zhang
*Beijing Electronic Science and Technology*
Beijing, China
zkj72@sina.com

5th Shuang Qing
*Beijing Electronic Science and Technology*
Beijing, China
shuangqling@126.com

6th Zhiqiang Wang*
*Beijing Electronic Science and Technology*
Beijing, China
wangzq@besti.edu.cn

*Abstract*—The braid group is an important non commutative group, at the same time, it is an important tool in quantum field theory with better topological structure, and often used as a research carrier for anti-quantum cryptographic algorithms. This paper proposed a difficult problem on a positive braid semi-group, and proved that the difficulty is not lower than the conjugate search problem. Based on this new difficult problem, we propose a new cryptosystem, which include a key exchange protocol and a public key encryption algorithm. Since our cryptosystem is implemented on a semi-group, it effectively avoids the analysis of attack algorithms on the cluster and makes our algorithm more secure.

## I. Introduction

The rapid development of quantum computers seriously threatens the security of classical public key cryptosystems. In 1994, Shor[1] proposed a quantum Fourier transform algorithm, which can construct an integer factorization polynomial (quantum) algorithm, which poses a substantial threat to the security of RSA. In 2003, Proos et al[2].extended the Shor algorithm to the elliptic curve, and obtained the polynomial (quantum) algorithm for solving the discrete logarithm problem over the elliptic curve, which poses a substantial threat to the security of ECC. Although quantum computation is still rapidly developing, almost all quantum algorithms can be uniformly attributed to the quantum Fourier transform on the commutative group. Moreover, these algorithms mainly attack the mathematics problems based on the commutative group, and they can't do anything about the non-commutative structure.

Braid group is an important non-commutative group, there are many difficult problems on it, which is suitable for designing cryptographic systems, and therefore, braid group is also used as a research carrier for anti-quantum cryptographic algorithms. Although some cryptographic algorithms on braid groups were attacked to varying degrees [3-7], according to research trends in the direction of anti-quantum attack cipher in recent years, the cryptographic scheme based on braid group is still hot, and new results have been put forward continuously [8,9,10].

Based on complex network environment, to prevent the attacks of quantum computers, this paper focuses on the research of the resistance quantum key exchange protocol. The classic key exchange protocol on the braid group is the AAG protocol [11], later, in order to resist the "length attack" [16], it was modified to AAFG protocol [12]. The modified AAFG agreement did not give too much expectation, soon afterwards, Lee SJ, Lee, E[13] and Dehornoy[14] pointed out that it was not safe enough. Diverse attack methods have caused many new schemes to be questioned quickly, resulting in almost no secure key exchange protocol on the braid group. Therefore, this paper will propose a new key exchange protocol to make up for this deficiency.

*Contributions*: All of the positive braids in the braid group contribute a semi-group, we propose a new computational problem on this semi-group, called exchange decomposition problem, it has been proved that the difficulty of this new problem is no less than the conjugate search problem on braid groups. Finally, based on the assumption of this problem, we propose a secure key exchange protocol.

## II. Preliminaries

### A. Definitions

Braid group is an important non-commutative group, there are many applications over it. In the following we will briefly review some important concepts related to the research of the paper. More details refer to [17].

**Definition 1** Define $B_n$ as a braid group generated by $a_1, a_2, \cdots a_{n-1}$, and if

$$\begin{cases} a_i a_j a_i = a_j a_i a_j & if \quad |i-j|=1 \quad (1) \\ a_i a_j = a_j a_i & if \quad |i-j| \geq 2 \quad (2) \end{cases}$$

A set $\Sigma = \{a_1, \cdots, a_2, a_{n-1}, a_1^{-1}, a_2^{-1}, \cdots a_{n-1}^{-1}\}$ is generated by the inverse of the generators and itself, which any sequence of it is a word on braid $B_n$.

**Definition 2** It is called positive braids, if a word does not contain its' any inverse. All of the positive braids form a semi-group, denoted as $B_n^+$. This is a very important semi-group, after the paper will describe a cryptographic algorithm in this semi-group.

*B. The fundamental braid and Normal form*

In braid groups, both of words is the same if consisting of braids in the same way or through the above two points (1),(2) can be transformed into the same form. So we need a unified standard form. This paper will use the left normal form [17]. Before introducing the standard type, we need to use the fundamental braid.

**Definition 3** Let $B_n$ be a braid group, the fundamental braid is defined as follow:

$$\begin{cases} \Delta_1 = e \\ \Delta_n = \Delta_{n-1} a_{n-1} a_{n-2} \cdots a_1 \end{cases}$$

From the geometric point of view, it is a special braid, $n$ braids intertwined with each other, but any two braids crossing only once. In case of confusion as $\Delta$, the fundamental braid has many good properties.

**Theorem 1** $\Delta_n^2 a_i = a_i \Delta_n^2$, $a_i \in B_n$

**Theorem 2** [11] if $A \leq \Delta^s$, then $\exists D_1, D_2 \in B_n^+$ s.t.

$\Delta^s = D_1 A = A D_2$.

If the braid group in each word only expressed as a kind of form, this form is called the standard type, the introduction of standard type can help us to distinguish each braids are equal. This paper introduces left normal form [17] proposed by Garside.

**Theorem 3**[17] Any $x \in B_n$ can be expressed as $x = \Delta^k y = \Delta^k s_1 \ldots s_r$, $k \in Z$, $s_i = \Delta \wedge (s_1 \ldots s_{i-1})^{-1} y$, we defined the word $x = \Delta^k s_1 \ldots s_r$ is $x$'s left normal form.

**Deduction** Any element $x \in B_n^+$, $\exists y \in B_n^+$, $m \in Z$, makes that $xy = \Delta^m$.

Proof: According to theorem 3, $x$ can be expressed in this form

$x = \Delta^k \overline{A} = \Delta^k s_1 \ldots s_r$, while $r \geq 0$, $x$ is a positive braid, it is easy to see that exist an element $y$ in $B_n^+$, s.t. $xy = \Delta$. While $r < 0$, $\overline{A}$ is a simple braid, we have $\overline{A} \leq \Delta$, then

$\exists D \in B_n^+$, s.t. $\overline{A} D = \Delta$

So whatever $n \geq -r-1$, for $n \geq 0$, and $y = \Delta^n D \in B_n^+$,

$$xy = \Delta^r \overline{A} \Delta^n D$$
$$= \Delta^{n+r} \Delta = \Delta^{n+r+1}$$

There are many difficult problems on the braid group, which can be used in the design of public key cryptosystems. The literature [19] summarizes the characteristics and conections of these issues in a more comprehensive way. Among them, the conjugate problem is a difficult problem which is widely studied. It is applicable to general non-exchange groups and is considered to be difficult to solve. The definition of the conjugate search problem as follow:

**Definition 4** (Conjugacy Search Problem) Given the values x and y, find a braid s in braid group $B_n$ such that

$$y = sxs^{-1}$$

III. A NEW hard problem

The design of public key cryptography is based on a difficult math problem. Now, we propose a new hard problem, called Exchange Decomposition Problem.

**Definition 5** (Exchange Decomposition Problem) Given the values U and V in semi-group $B_n^+$, find two positive braids s and t such that :

$$\begin{cases} st = U \\ ts = V \end{cases}$$

We will prove that the Exchange Decomposition Problem is at least as hard as the conjugate search problem.

Based on the above two hard problems(Conjugate Search Problem and Exchange Decomposition Problem), we propose the following two security assumptions:

**Assumption1** We assume that it is hard to compute s,t, given the values U and V in Bn, where st = U and ts = V.

**Assumption2** We assume that it is hard to compute s, given the values x and y, where $y = sxs^{-1}$.

**Theorem 4** Assumption1 holds if and only if assumption 2 holds.

Proof

Assume that an adversary A attacks the assumption1, while another adversary B attacks the assumption2. What we need to do is to prove that

(1) If A attacks successfully, then B attacks successfully.

(2) If B attacks successfully, then A attacks successfully.

For (1), if A attacks successfully, then the t and s can be computed by the given the values U and V. Assume that

$$xs^{-1} = c, \text{ then } y = sc, x = cs$$

So we can compute s easily.

For (2), if B attacks successfully, then the s can be computed by given the values x and y. Consider that

$$\begin{cases} st = U \\ ts = V \end{cases}$$

We can see

$$tst = tU = Vt, \text{ then } V = tUt^{-1}$$

$$sts = Us = sV, \text{ then } U = sVs^{-1}$$

So we can compute $s$ and $t$ easily.

In summary, the theorem is proved.

The above has proved that difficulties of the two problems are the same on the braid group $B_n$, but there is no inverse of any element in a semi-group $B_n^+$. Therefore, based on this new computing problem, we can put forward a better cryptographic algorithm in a semi-group. Next section, we will propose a cryptographic algorithm on positive braid semi-groups.

IV. KEY EXCHANGE PROTOCOL

The key exchange protocol is as follows: there are two parties, called Alice and Bob. They would like to agree on a common secret in such a way that an adversary observing the median values of communication cannot deduce any useful information about the secret they shared.

A. AAG protocol

The public key consists of two sets of braids in $B_n$. The secret key of Alice is a word $s$ and its inverses generated by a random function $u$, while the secret key of Bob is a word $t$ and its inverses generated by a random function $v$. The scheme is as follows [11]:

---

- A computes the braid $s = u(p_1, ..., p_l)$, sends $q_i = sq_is^{-1}$, $i = 1, ..., m$ to B;

- B computes the braid $r = u(q_1, ..., q_m)$, sends $p_i = rp_ir^{-1}$, $i = 1, ..., l$ to A;

- A computes $t_A = s \cdot u(p_1', \cdots, p_l')$;

- B computes $t_B = v(q_1', \cdots, q_m') \cdot r^{-1}$.

The common key is $t_A = t_B$.

---

It is easy to see that

$$\begin{aligned} t_A &= s \cdot u(p_1', \cdots, p_l') \\ &= s \cdot r \cdot u(p_1, \cdots, p_l) r^{-1} \\ &= srs^{-1}r^{-1} = s \cdot v(q_1, \cdots, q_m) \cdot s^{-1} \cdot r^{-1} \\ &= v(q_1', \cdots, q_m') \cdot r^{-1} = t_B \end{aligned}$$

The security is based on the difficulty of a variant to Conjugate Search Problem in $B_n$, namely the Multiple Conjugate Search Problem, in which one tries to find a conjugating braid starting not from one single pair of conjugate braids (p, p'), but from a finite family of such pairs $(p_i, p_i')$ obtained using the same conjugating braid [18].

B. A new key exchange protocol

Choose two sets of random braids

$$<p_1, p_2, ..., p_l>, \text{ and } <q_1, q_2, ..., q_m>$$

As the public key of Alice and Bob on semi-group $B_n^+$. Alice and Bob use their own public key as parameters, then get their own private keys by the private key generators $u$, $v$, and private key generation is confidential. $k$ and $h$ are open and large enough to meet the experimental conditions. Perform the following steps:

**Step 1:** Alice computes the positive braids $s = u(p_1, ..., p_l)$, computes $s_1 \in B_n^+$, where $ss_1 = \Delta^{2k}$, $k$ is a public integer. Then, use $s, s_1$ to compute

$$q_i' = sq_is_1, \ i = 1, \cdots, m$$

and send $<q_1', q_2', ..., q_m'>$ to Bob.

**Step 2:** Bob computes the positive braid $r = v(p_1, ..., p_l)$, computes $r_1 \in B_n^+$, where $rr_1 = \Delta^{2h}$, $h$ is a public integer. Then, use $r, r_1$ to compute

$$p_i' = r_1p_ir, \ i = 1, \cdots l.$$

and send $<p_1', p_2', ..., p_l'>$ to Alice.

**Step 3:** Alice computes $S = u(p_1', p_2', ..., p_l') \cdot s_1 / \Delta^{2k(l-1)}$.

**Step 4:** Bob computes $T = r_1 \cdot v(q_1', q_2', ..., q_m') / \Delta^{2h(m-1)}$.

It is easy to see

$$\begin{aligned} S &= u(p_1', p_2', ..., p_l') \cdot s_1 / \Delta^{2k(l-1)} \\ &= \Delta^{2k(l-1)} r_1sr \cdot s_1 / \Delta^{2k(l-1)} \\ &= r_1srs_1 \end{aligned}$$

$$\begin{aligned} T &= r_1 \cdot v(q_1', q_2', ..., q_m') / \Delta^{2h(m-1)} \\ &= r_1 \cdot \Delta^{2h(m-1)} srs_1 / \Delta^{2h(m-1)} \\ &= r_1srs_1 \end{aligned}$$

∴ S = T

This shows that the process allows Alice and Bob to obtain a common secret, $S = T$, and distribute the key safely, and enable users to create shared symmetric keys.

In formal terms, our agreement does not seem to be very different from the AAG protocol. In fact, they are essentially different.

| | Group | Hard Problem |
|---|---|---|
| AAG protocol | Braid group | Multiple Conjugate Search Problem |
| Our protocol | Semi-group | Exchange Decomposition Problem |

All attack algorithms for AAG protocol are running on the group algorithm. However, our protocol is based on the semi-group generated by the braid group $B_n$. In addition, it has been proved that the new problem which applied by the protocol is at least as hard as the conjugate search problem on the braid group. All of the elements in the semi-group are positive braids, there is no inverse, so it will resist more attacks.

## V. PUBLIC KEY ENCRYPTION ALGORITHM

In the previous chapter, we proposed a new key exchange protocol that compares it with the AAG protocol and also makes a simple analysis of the protocol. By using the key exchange protocol by previous chapter, we construct a new public key encryption algorithm. Let $H : B_n \to \{0,1\}^k$ be an ideal that has function from the braid group to the message space.

**1. Key Generation:**

(a) Choose two sets sufficiently complicated braids

$< p_1, p_2, ..., p_l >$, $< q_1, q_2, ..., q_m > \in B_n$

(b) Let $d$ be a sufficiently complicated integer, choose a random function $u$, compute $s = u(p_1, p_2, ..., p_l)$, $s_1$, where $ss_1 = \Delta^{2d}$, then compute $q_i' = sq_i s_1$, $i = 1,...,m$.

(c) The public key $pk$ is

$(d, < q_1', \cdots, q_m' >, < q_1, \cdots, q_m >)$,

where $q_i' = sq_i s_1$, $i = 1,...,m$;

The secret key $sk$ is $(s, s_1)$.

**2. Encryption:** Given a message $m \in \{0,1\}^k$ and $pk$.

(a) Choose a random function $v$, compute

$r = v(q_1, \cdots, q_m)$, $r_1$, where $rr_1 = \Delta^{2t}$, then compute $p_i'$, where $p_i' = r_1 p_i r, i = 1, \cdots, l$;

(b) Set $Y = p_1' p_2' | \cdots | p_l'$, $Z = r_1 v(q_1', \cdots, q_m')/\Delta^{2d(m-1)}$;

(c) Compute $k = H(Y, Z)$, $c = E_k(m)$, the cipher text

$(Y, c)$.

**3. Decryption:** Given a cipher text $(Y, c)$ and $sk$, compute

$Z = u(p_1', \cdots, p_l') \cdot s_1 / \Delta^{2t(l-1)}$, $k = H(Y, Z)$, $m = D_k(c)$.

Compared to the PKC proposed by Ko et al.[19], our algorithm does not need to construct a commutative subgroup. Because of the non-commutativity of the positive braids, our encryption algorithm can better resist the quantum Fourier's attack.

## VI. CONCLUSION

In this paper, firstly, we introduced the research status of the cryptosystem based on the braid group and the basic concepts of the braid group. Then, a new computational problem was proposed, and proved that the problem is at least as hard as the conjugate search problem over the braid group. Finally, we proposed a new key exchange protocol and construct a new public key encryption. In fact, our cryptosystem is implemented in a semi-group, which makes many of the attacks in the past will no longer work. Therefore, the new cryptosystem is more secure.

ACKNOWLEDGMENT

This work was supported by National key R & D plan under grand :No. 2017YFB0801803.